\documentclass[12pt]{article}
\usepackage{amsmath, amssymb}
\begin{document}

\title{
CONCERNING BJORKEN'S MODEL\\
OF\\
SPONTANEOUS BREAKDOWN OF LORENTZ INVARIANCE}

\author{R. Acharya\\
Department of Physics \& Astronomy\\
Arizona State University\\
Tempe, AZ 85287}

\date{February 2006}
\maketitle

\begin{abstract}

We revisit Bjorken's model of spontaneous breakdown of Lorentz
invariance. We show that the model possesses zero mass, spin zero
(scalar) Nambu-Goldstone boson, in addition to the zero mass, spin
one (vector) photon.
\end{abstract}

Following the pioneering work of Nambu and Jona-Lasinio (NJL)
\cite{Nambu} on the spontaneous breaking of
global chiral symmetry $\psi \rightarrow \exp (i \alpha \gamma_{5})
\psi$ which demonstrated that massless pions could emerge as
composite Nambu-Goldstone (NG) bosons, in a theory whose
fundamental fields were spin $\tfrac{1}{2}$ fermions interacting
via four-fermion interaction, Bjorken \cite{Bjorken1} proposed a
similar idea for what he termed as the ``dynamical generation of
quantum electrodynamics" (QED) without invoking \underline{local}
$U(1)$ gange invariance as a fundamental axiom. In Bjorken's
model, the masslessness of the photon is attributed to the
spontaneous breakdown of Lorentz invariance with a vector
four-fermion interaction. The Lagrangian in Bjorken's model is
given by

\begin{equation}\label{1}
L = \overline{\psi} (\gamma^{\mu}\partial_{\mu} - m) \psi - \lambda
(\overline{\psi} \gamma^{\mu} \psi) (\overline{\psi} \gamma^{\mu} \psi)
\end{equation}

We observe that Eq.~(\ref{1}) retains \underline{global} $U(1)$
invariance, i.e., under $\psi \rightarrow \exp (i \alpha) \psi$.
Bjorken \cite{Bjorken2} revisited his 1963 work recently. Kraus
and Tomboulis \cite{Kraus} have resurrected Bjorken's idea to
explore whether photons \underline{and} gravitons could arise as
spontaneous breakdown of Lorentz invariance. Jenkins has further
explored this idea, which provides an avenue to sidestep the
Weinberg-Witten theorem i.e., a Lorentz invariant theory with a
Lorentz invariant vacuum state and a Lorentz covariant
energy-momentum tensor does not permit a composite graviton
\cite{Jenkins}. Kostelecky and collaborators \cite{Bluhm} have
presented a general treatment of spontaneous local Lorentz and
diffeomorphism violation for Minkowski and Riemann spacetimes. It
is pertinent to recall here that the validity of Goldstone's
theorem as first spelled out in the classic 1962 paper of
Goldstone, Salam and Weinberg \cite{Weinberg1} crucially depends on
a manifest covariance of the underlying local theory, i.e., a
theory in which $\partial_{\mu}j^{\mu} = 0$ and the vanishing of
of surface terms at infinity. This then leads to the result that
$Q_{V} \equiv \int \limits_{V} d^{3} x j^{0}(x)$ has a zero mass,
zero spin boson in its spectrum.

We briefly highlight the salient points of Bjorken
model, Eq.~(\ref{1}), following Guralnik \cite{Guralnik}.

The ``boundary conditions" imposed on Eq.~(\ref{1}) are

\begin{equation}\label{2}
\langle j^{\mu} \rangle_{0} = n^{\mu} \neq 0
\end{equation}

where $j^{\mu} = \overline{\psi} \gamma^{\mu} \psi$.

The commutator of $j^{\mu}$ with the generator of Lorentz
transformations $M^{\mu \nu}$ is given by

\begin{equation}\label{3}
[ M^{\mu \nu}, j^{\lambda} ] = g^{\mu \lambda} j^{\nu} - g^{\nu
\lambda} j^{\mu}.
\end{equation}

From Eq.~(\ref{2}) and Eq.~(\ref{3}), we conclude that

\begin{equation}\label{4}
M^{\mu \nu} \quad |_{0} \rangle \neq 0
\end{equation}

i.e., Lorentz invariance of the vacuum state is broken and
$j^{\mu} |_{0} \rangle$ must contain a zero mass, spin one
particle, i.e., a photon.

In this brief note, our chief concern is the recent
striking assertion in a series of papers by Arkani-Hamed et. al
\cite{Arkani} that ``we cannot have spontaneously broken Lorentz
invariance without having additional dynamic effects. The reason
is that there \underline{must} be Goldstone bosons (scalar)
associated with the (spontaneous) breaking of Lorentz invariance."

We shall demonstrate the validity of the above-mentioned assertion
by going back to Bjorken's 1963 model, Eq.~(\ref{1}). The argument
is straightforward. We recall that if the vacuum expectation
value of the vector current, $ j^{\mu} = \overline{\psi} \gamma^{\mu} \psi$
were to vanish, i.e., $\langle j^{\mu} \rangle_{0} = 0 $
(and hence \underline{No} spontaneous breakdown of Lorentz
invariance), then the charge (Lorentz scalar) $Q_{V} \equiv \int
\limits_{V} d^{3} x j^{0}(x)$ must annihilate the vacuum, $Q_{V}
|_{0} \rangle = 0$ \cite{Weinberg2}. On the contrary, if the vacuum
spontaneously breaks Lorentz invariance, $\langle j^{\mu}
\rangle_{0} \neq 0$, then $Q_{V}$ cannot annihilate the vacuum: $Q_{V}
|_{0} \rangle \neq 0$ and a spontaneous broken global symmetry is
always associated with a degeneracy of the vacuum state.
Therefore, the spontaneous breakdown of Lorentz invariance,
``triggers" the spontaneous breakdown of global $U(1)$ symmetry,
exhibited by the Bjorken model, Eq.~(\ref{1}) with a conserved
current $\partial_{\mu} j^{\mu} = 0$ ($j^{\mu} = \overline{\psi}
\gamma^{\mu} \psi$) obeying local commutativity, i.e., $[j^{\mu}(x),
j^{\mu}(y)] = 0$ for space-like separations. Since Eq.~(\ref{1})
retains the global $U(1)$ invariance of the free Lagrangian, we
immediately conclude that the Bjorken model must have (composite)
scalar NG boson, in addition to the usually recognized (composite)
zero mass vector photon. The presence of the scalar NG boson seems
to have escaped attention. In conclusion, the conjecture of
Arkani-Hamed and collaborators \cite{Arkani} is verified and
justified in the present context of the Bjorken model,
Eq.~(\ref{1}).


\begin{thebibliography}{99}

\bibitem{Nambu}
Y. Nambu and G. Jona-Lasinio, \emph{Phys. Rev.} \textbf{122}, 345 (1961).

\bibitem{Bjorken1}
J. Bjorken, \emph{Annals Phys.} \textbf{24}, 174 (1963).

\bibitem{Bjorken2}
J. Bjorken, ArXiv: hep-th/0111196.

\bibitem{Kraus}
P. Kraus and E. Tomboulis, \emph{Phys. Rev. D} \textbf{66},
045015 (2002).

\bibitem{Jenkins}
A. Jenkins, \emph{Phys. Rev. D} \textbf{69}, 105007 (2004);\\
S. Weinberg and E. Witten, \emph{Phys. Lett. B} \textbf{96}, 59
(1980).

\bibitem{Bluhm}
R. Bluhm and V. Kotelecky, ArXiv: hep-th/0412320.

\bibitem{Weinberg1}
S. Weinberg, \emph{Quantum Theory of Fields,} VII, Cambridge
University Press (1996).

\bibitem{Guralnik}
G. Guralnik, \emph{Phys. Rev.} \textbf{136} B1404 (1964).

\bibitem{Arkani}
N. Arkani-Hamed et.al., ArXiv:hep-ph/0407034; hep-ph/0507120 and
references therein.

\bibitem{Weinberg2}
S. Weinberg, \emph{1970 Brandeis University Lectures on Elementary
Particles and Quantum Field Theory,} Vol. 1, p. 301, MIT Press
(1970).

\end{thebibliography}
\end{document}